# Tuning One-Dimensional Plasmonic Gap at Nanometer Scale for Advanced SERS Detection

*Mahsa Haddadi Moghaddam, Sobhagyam Sharma, Daehwan Park, Dai Sik Kim\**

A. B.  Mahsa Haddadi Moghaddam, Sobhagyam Sharma, Daehwan Park.

A. B. C. D.  Dai Sik Kim, E-mail: daisikkim@unist.ac.kr

A- Department of Physics, Ulsan National Institute of Science and Technology, Ulsan 44919, Republic of Korea

B- Quantum Photonics Institute, Ulsan National University of Science and Technology, Ulsan 44919, Republic of Korea

C- Center for Angstrom Scale Electromagnetism, Ulsan National University of Science and Technology, Ulsan 44919, Republic of Korea

D- Department of Physics and Astronomy, Seoul National University, Seoul 08826, Republic of Korea

\*Mahsa Haddadi Moghaddam and Sobhagyam Sharma contributed equally.

## Abstract

The "hotspots", which are typically found in nanogaps between metal structures, are critical for the enhancement of the electromagnetic field. Surface-enhanced Raman scattering (SERS), a technique known for its exceptional sensitivity and molecular detection capability, relies on the creation of these hotspots within nanostructures, where localized surface plasmon resonance (LSPR) amplifies Raman signals. However, creating adjustable nanogaps on a large scale remains challenging, particularly for applications involving biomacromolecules of various sizes. The development of tunable plasmonic nanostructures on flexible substrates represents a significant advance in the creation and precise control of these hotspots.

Our work introduces tunable nanogaps on flexible substrates, utilizing thermally responsive materials to allow real-time control of gap width for different molecule sizes. Through advanced nanofabrication techniques, we have achieved uniform, tunable nanogaps over large areas wafer scale, enabling dynamic modulation of SERS signals. This approach resulted in an enhancement factor of over ~$10^7$, sufficient for single-molecule detection, with a detection limit as low as $10^{-12}$ M. Our thermally tunable nanogaps provide a powerful tool for precise detection of molecules and offer significant advantages for a wide range of sensing and analytical applications.

## 1. Introduction

The advancement of chemical and biochemical analysis techniques has revolutionized the ability to detect and identify molecules with high precision. Among these techniques, surface enhanced Raman scattering (SERS) stands out as a transformative tool due to its unparalleled sensitivity and specificity.[1-9] This method amplifies Raman signals by localized surface plasmon resonance (LSPR) in metal nanostructures and enables the detection of molecules with single-molecule sensitivity without damaging the samples.[10-13] A critical component of SERS is the creation of "hotspots" regions within the nanostructures where the electromagnetic (EM) fields are significantly amplified .[14-17] These hotspots are typically located in the nanogaps between metallic structures, with the width of the gap playing a crucial role in determining the strength of the electromagnetic field. Narrower gaps lead to stronger local fields and thus to an amplification of the Raman signal.[18-24] One of the most promising advances in SERS technology is the ability to adjust the width of the nanogaps to the target molecules. This capability is particularly important for applications involving biomacromolecules, which can vary greatly in size. Zhao et al. introduced a cost-effective method for fabricating sub-10 nm metallic nanogaps using swelling-induced nano cracking.[25] Ye et al. prepared periodic folded Au nanostructures using atomic force microscopy (AFM) scratching and nano skiving.[26] Im et al. demonstrated sub-10-nm metal nanogap arrays, in which vertically aligned plasmonic nanogaps are formed between two metal structures by a sacrificial layer of ultrathin aluminum oxide grown by atomic layer deposition.[27]

Plasmonic bowtie nanostructures were precisely constructed with Zhan et al. using a DNA origami-based bottom-up assembly strategy, allowing control over the bowtie's geometry with an approximate 5 nm gap, and a single Raman probe was accurately positioned at this gap.[24]

Although current fabrication methods offer excellent scalability for the production of nanogaps, challenges remain due to the accumulation of molecules above the gaps and the difficulty of direct insertion. Additionally, fabricating multiple samples to achieve different gap sizes tailored to different molecules is time-consuming and resource intensive. To overcome this problem, tunable nanogaps can precisely adjust the size of the gaps in the nanometer scale, allowing for efficient customization for different molecular applications.[28-34] This adaptability is essential for detecting small dye molecules, which require extremely narrow nanogaps, as well as larger biomacromolecules that benefit from wider gaps to effectively reach the hotspots.[35-37]

In this work, we have successfully developed tunable nanogaps on flexible substrates using thermally responsive materials that allow precise control of the gap width to accommodate different molecule sizes and ensure optimal interaction and signal amplification.[33, 38-39] Using advanced nanofabrication techniques, such as photolithographic patterning of metals and transfer technology, we have created periodic one-dimensional (1D) uniform nano-gaps on flexible substrates over large areas at wafer scale. These nanogaps can be finely tuned to dimensions below 100 nm, especially for smaller periodicities, demonstrating exceptional control over the structural parameters. We have shown that by manipulating the temperature, both long-wavelength transmission and surface-enhanced Raman scattering (SERS) can be dynamically tuned in real time. The ability to precisely control the gap width through temperature variations allowed us to achieve a significant overlap between the localized surface plasmon resonance and the vibrational energy of the Raman modes in the molecules trapped within the nanogap. This resonance overlap leads to a significant amplification of the SERS signal, resulting in an enhancement factor (EF) of over $\sim 10^7$, which is sufficient for single molecule detection. In addition, our thermally tunable nanogaps exhibit extremely high SERS sensitivity, with a demonstrated detection limit of only $10^{-12}$ M. This high sensitivity underscores the effectiveness of our approach and makes it a powerful tool for applications requiring precise molecular detection and analysis. The ability to dynamically adjust the dimensions of the nanogaps and amplify the SERS signals through temperature control offers significant advantages for a wide range of sensing and analysis applications.

## 2. Results and Discussion

### 2.1. Fabrication and characterization of thermally tunable nanogaps

To achieve plasmonic permutations, we used a nanogap device based on a thermal tunable substrate as a template for anchoring R6G molecules into the nanogap by employing deep coating and rinsing methods (**Figure 1**a). This experimental setup significantly enhances the Raman scattering of the molecules due to the intense field confinement within the nanogap. In contrast to tip-enhanced approaches, the continuous 1D structure of the nanogap enables large-area Raman sensing and provides spectral tunability in SERS detection.[40] In addition, the strong temperature sensitivity of the PDMS substrate in relation to the Au metal enables precise tuning of the gap size to the target molecules by simply varying the temperature. Figure 1b illustrates the transfer process of the nanogap to the PDMS substrate. The nanogap patterns with different periodicities (P) are first formed by the photolithography method on a sacrificial layer deposited on a rigid substrate, followed by the deposition of the thin Au films with a uniform thickness of 50 nm, consecutively (For more details, see the Supporting Information **S1**). Subsequently, A thin and uniform PDMS film is spin-coated onto the aligned nanogaps and then cured. To improve both flexibility and performance, 3-mercaptopropyltrimethoxysilane (MPTMS) is applied as a self-assembled monolayer between the Au patterns and the PDMS layer. Finally, the Au nanogap film is transferred to the PDMS substrate by chemical etching of the underlying sacrificial layer, allowing easy separation from the rigid substrate.[33] PDMS is a good platform to improve temperature sensitivity due to its unique material properties.[41-43] The thermal expansion coefficient of PDMS ($310 \times 10^{-6}$ °$C^{-1}$), is much higher than that of metallic materials, therefore, the temperature sensitivity can be effectively improved by the tensile force generated by the thermally expanded PDMS and this led to opening and closing the gap between two gold slits (Figure 1d). Figure 1c shows the field emission scanning electron microscopy (FE-SEM) images of the nanogap with a periodicity of 5 μm in the relaxed state with a closed gap, while Figure 1d shows the same nanogap in the stretched state with a gap width of about 130 nm (the insets show the magnification of the single gap). When the PDMS substrate expands with increasing temperature, the gold layer of the trenches no longer has too much effect (due to its low coefficient of thermal expansion of $14 \times 10^{-6}$ °$C^{-1}$), which leads to the opening of the nanogaps.[44] Since PDMS is a chemically stable elastomeric polymer, the thermally changing gap in our structure returns to the original position without dimensional stability problems. In addition to the high coefficient of thermal expansion, the design of thermomechanical devices using PDMS requires a thorough understanding of its temperature-dependent mechanical properties, such as Young's modulus.[45] Studies have shown that increasing the temperature of PDMS leads to a decrease in the elastic modulus, while the Poisson's ratio remains largely unchanged.[41] The lower modulus of elasticity, which translates into lower stiffness, makes PDMS an ideal material for applications that require high deformation by temperature modulation and offer significant mechanical flexibility.[46]

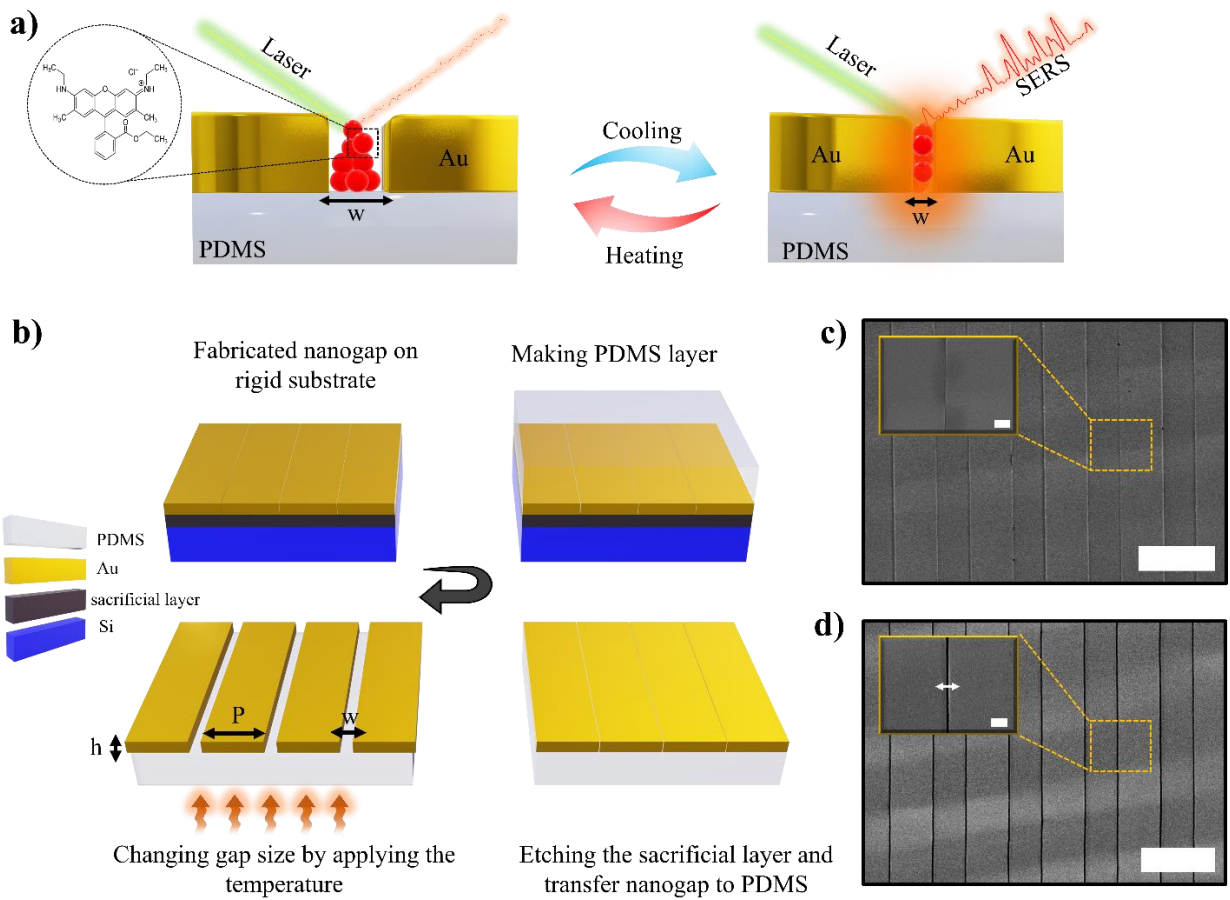

**Figure 1.** a) Schematic representation of the SERS sensing mechanism, showing molecules trapped in the nanogap. b) Schematic diagram of the transfer process of nanogaps with micrometer periodicity onto a flexible substrate. FE-SEM images of the top view of the nanogap c) before and d) after strain (scale bars: 10 μm). (The inset shows a close-up of a single gap from top, (scale: 1 μm).

## 2.2. Optical and electrical properties of thermally tunable nanogaps

We performed broadband spectroscopy in the microwave region and electrical measurements under different temperature conditions to evaluate the sensitivity of the samples to the gap opening. Transmission measurements of the nanogaps with different periodicities were performed in Ku-band (12−18 GHz) using a vector network analyzer while the temperature gradually increased in 10-degree steps (see **Figure S2**, Supporting Information). **Figure 2**a shows the normalized transmission amplitude for the nanogaps with a periodicity of 5 µm as the temperature increases from room temperature to 110 °C. The transmission amplitude exceeds 50 % at 110 °C, which indicates a significant gap opening at high temperatures. In the previous work, we have demonstrated that the periodicity of the gap of the plasmonic nanostructures has a great influence on the uniformity of the hotspots and the intensity of the electromagnetic field.[33] Therefore, the relationship between the periodicity and the distribution of the electric field is investigated in detail. Figure 2b shows the transmission sensitivity at 15 GHz for different periodicities plotted against temperature. The gap opening becomes more sensitive to lower stress with increasing periodicity (e.g. up to 50 µm), so that even a small change in temperature leads to a significant gap enlargement. These results suggest that using smaller periodicities provides the best precision for controlling the gap size in the nanometer range. Figure 2c shows the hysteresis loops for the 5 µm nanogap during cyclic stretching and relaxation at 15 GHz caused by alternating temperature changes due to heating and cooling of the sample. The hysteresis loops for the 10 µm and 25 µm periodicities under identical conditions are included in the supporting information **Figure S3**a and S3b, respectively. The performance of the thermally tunable nanogaps is further evaluated by their relative resistance at different temperatures. As shown in Figure 2d, analysis of the samples with different periodicities shows that the 1D nanogap has a wide sensitivity range, with resistance values varying from 5 Ω to 160 Ω as the temperature increases from 23 °C to 110 °C. The increasing temperature leads to an external stress that progressively increases the average resistance between neighboring gold bars as the electrical contact decreases. Consistent with Figure 2b, the change in resistance is more pronounced at larger periodicities. Figure 2e illustrates the stepwise change in resistance with a temperature increase from room temperature to 110 °C in 10-degree steps for a 5 μm periodicity. By comparing these resistance changes with the gap widening observed in previous SEM data, we found that the gap expands to approximately 230 nm at 110 °C. **Figure S4** shows the stepwise gap opening with increasing temperature for 10 μm and 25 µm periodicity, showing gap expansions of 350 nm and 510 nm, respectively. Figure **S4** also shows the corresponding cooling procedure to the room temperature.

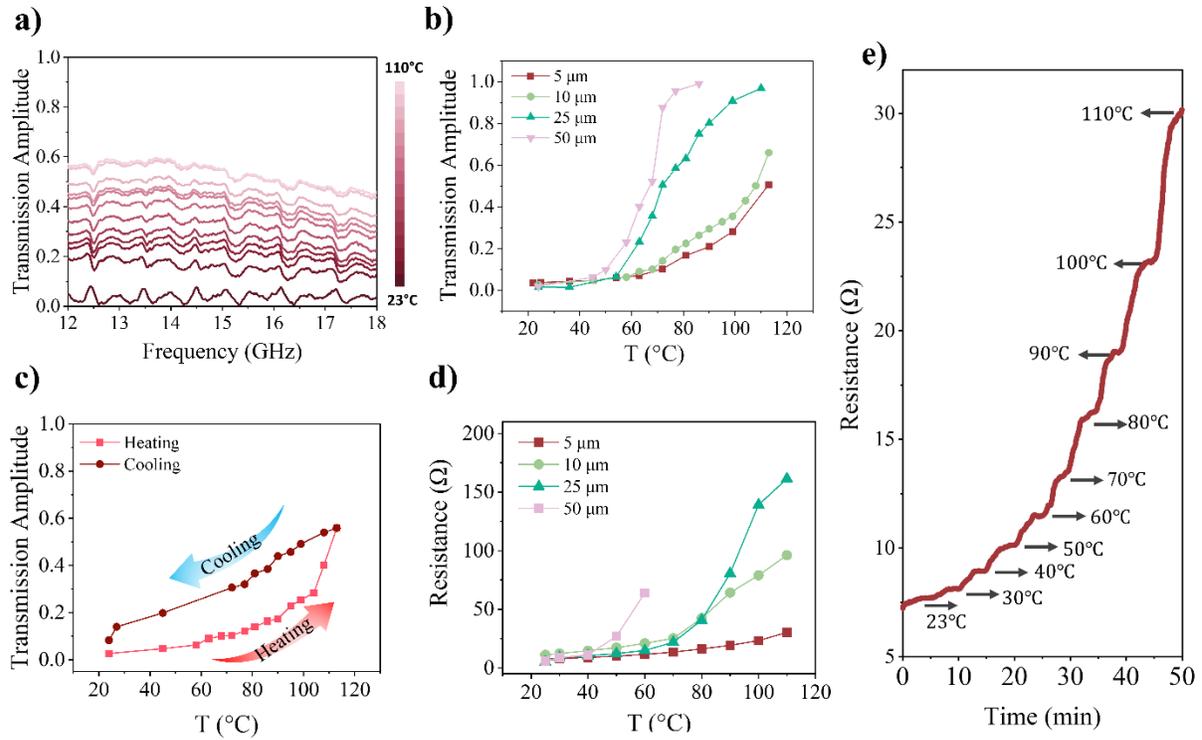

**Figure 2.** a) Normalized transmission spectra of a nanogap with 5 μm periodicity under various temperature in the microwave frequency range of 12 to 18 GHz. b) Amplitude transmission variations at 15 GHz as a function of temperature, for different periodicities. c) Hysteresis loops of the 5 μm periodicity gap during cooling and heating cycles at 15 GHz. d) Relative resistance against temperature for different periodicities. e) Stepwise resistance changes with temperature with steps of 10 degrees.

## 2.3. Numerical Simulations of electric Field Enhancement of nanogap

Numerical simulations were performed to gain a deeper understanding of the effects of the dynamic contraction process on the enhancement of the electric field within the 1D nanogap. To quantitatively estimate the variation of the gap width with temperature, we modeled the Au nanogap on the PDMS substrate and calculated the gap width ($w_{gap}$) as a function of temperature with a periodicity of 5 µm (see Supporting Information **Figure S5**). The simulations revealed that the gap widens by less than 30 nm for every 10-degree increase in temperature. This demonstrates the precise control achieved through temperature adjustments, which is a challenge for sub-10 nm gaps widening using mechanical strain alone.[33] The COMSOL simulations and experimental results of gap opening, derived from top-view SEM images, demonstrate strong agreement with respect to temperature (**Figure 3**a). Figure 3b shows the cross-sectional profile of the distribution of the optical field ($\lambda$ = 532 nm) of the simulated model for different gap sizes, with a constant gold thickness of 50 nm to replicate the same conditions. The 10 nm nanogap shows strong near-field coupling, which significantly enhances the electric field due to the coupling of the localized surface plasmon resonance (LSPR) between adjacent gold surfaces.[47] However, as the gap width increases from 10 to 100 nm, only a slight increase in electric field strength is observed at the gap edges, indicating weaker near field coupling for large gaps. The enhancement factor of the nanogap, denoted as $EF_{gap}$, can be estimated by $EF_{gap} = |E/E_0|^4$, where E represents the magnitude of the enhanced optical field at the nanogap and $E_0$ is the magnitude of the optical excitation field (see Figure 3c).[19, 25] For 1D nanogap arrays, simulations indicate that the total SERS enhancement factor for the active region under a diffraction-limited focused beam spot, reaches $0.7 \times 10^9$ for a 1 µm long nanogap with a gap width of 10 nm (see Supporting Information **S6** for details). Figure 3d illustrates the distribution of the electric field ($E/E_0$) within the nanogap, spanning from the bottom (substrate) to the edge, for gap size ranging from 5 to 50 nm. The results show that maximum enhancement is achieved when the target molecules are localized within the nanogap, particularly near its edges. Figure 3e presents the simulation results of the electric field distribution inside 10 nm gaps with different depths. The enhancement factor of the nanogap peaks at the gap depth of 50 nm, (Figure 3f) facilitating a stronger field enhancement at smaller gap depth. The results also indicate that the electric field at the non-metallic bottom (here PDMS) of the nanogaps is relatively weak, attributed to energy dissipation during transmission. Finally, we show that the thickness of the substrate has only a minimal effect on the enhancement of the electric field (see Supporting Information, **Figure S7**).

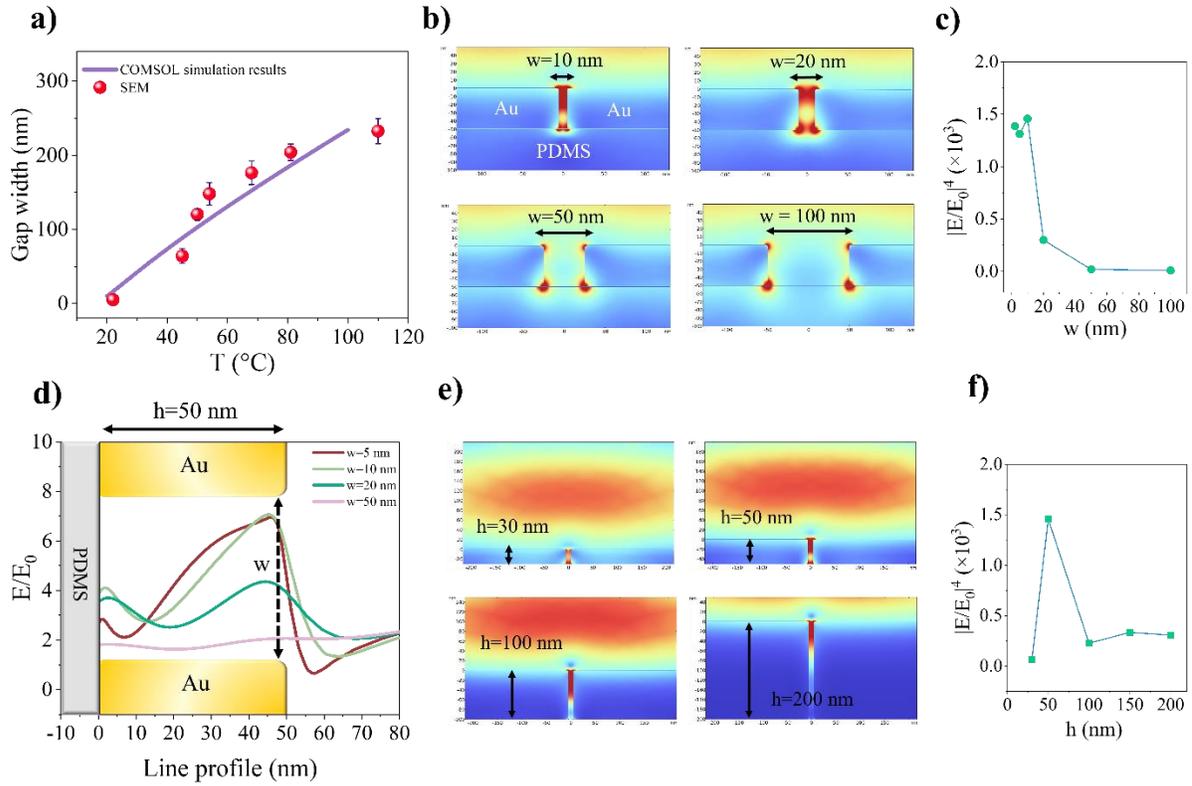

**Figure 3.** a) Comparison of COMSOL simulations and experimental data for the gap opening as a function of temperature. b) COMSOL-simulated electric field ($|E/E_0|$) distribution at 532 nm excitation for nanogaps with a 50 nm depth and varying gap widths. c) Calculated field enhancement $|E/E_0|^4$ for nanogaps with a depth of 50 nm and different gap widths. d) Electric field ($E/E_0$) distribution across the nanogap, from the substrate to the edge. e) COMSOL-simulated electric field ($|E/E_0|$) distribution at 532 nm excitation for nanogaps with a 10 nm width and varying gap depths. f) Calculated $|E/E_0|^4$ values for nanogaps with different depths and a fixed gap width of 10 nm.

## 2.4. SERS Spectral Investigations of thermally tunable nanogaps

We have further investigated the effects of localized surface plasmon resonance (LSPR) of nanogaps by analyzing their performance in surface-enhanced Raman scattering (SERS) for the detection of molecules. Our observations show a clear dependence of the Raman intensity on the gap width. **Figure 4**a and 4b illustrate the schematic representations of the gap contraction during the cooling process and the gap expansion during the heating process of the substrate, respectively, and demonstrate how molecules can be effectively confined within the nanogap. In this study, a sample with a one-dimensional (1D) gold pattern on a PDMS substrate with 5 μm periodic nanogaps is used as a SERS probe, with rhodamine 6G (R6G) molecules serving as the Raman probe. Each nanogap within a single array act as a hotspot for Raman measurement, with an effective area of approximately 1 μm based on the laser spot size. Precise control of the gap is critical for molecular detection, and the 5 μm periodicity allows fine tuning by adjusting the temperature. Based on our simulation results, the Au layer thickness was optimized to 50 nm to achieve optimal SERS performance. The incident light used is unpolarized with a wavelength of 532 nm, and the SERS spectra are collected over the range of 700 to 1800 $cm^{-1}$. As demonstrated in **Figure S8**a in the Supporting Information, the intensity of the characteristic Raman peaks increases when the temperature decreases from 70 °C to -20 °C. This enhancement is due to the narrowing of the nanogap width caused by the contraction of the PDMS during the cooling process and the adaptation to the molecules within the nanogaps. Conversely, Figure S8b shows the heating process and demonstrates the reverse effect, where the nano-gap widens when the temperature returns to its initial state.

We performed a comparative analysis between the thin film and the 1D nanogap, ensuring that all experiments and measurements were performed under identical conditions. **Figure S9** in the Supporting Information shows the differences in Raman intensity of the thin film at different temperatures. In contrast to the 1D nanogap structure, the thin film did not show a consistent trend in Raman intensity with decreasing temperature, indicating that the high Raman intensity comes from the change in the size of the nanogaps. Afterwards, Raman spectra of $10^{-4}$ M R6G were collected from a random spot on the 1D closed gap at -10 °C, the nanogaps at room temperature, and from the Au film of the same thickness on the PDMS substrate (Figure 4c). The closed gap at -10 °C exhibited a significantly enhanced Raman signal for characteristic peaks compared to the other conditions, highlighting its superior performance in fitting the molecules inside the gaps. From this it can be deduced that the size of the gap depends on the thickness of the enclosed molecules, which indicates that the dimensions of the gap are directly influenced by the thickness of the molecule. Figure 4d illustrate the analytical enhancement factor calculated at Raman shifts of 771 cm$^{-1}$, 1184 cm$^{-1}$, 1359 cm$^{-1}$, 1505 cm$^{-1}$, and 1646 cm$^{-1}$ across different temperatures. Results indicate that most of the characteristic peaks reach saturation at -10 °C, indicating that the nanogap closes at this temperature according to the molecule size. The significant enhancement factor of $0.53 \times 10^7$ at 1359 cm$^{-1}$ emphasizes the strong electromagnetic enhancement within the 1D nanogap, with optimal coupling to the molecular vibrations (see Section **S10** in the Supporting Information for more details).[9, 48-49] Subsequently, the Raman spectra of R6G molecules with different concentrations ranging from $10^{-3}$ M to $10^{-12}$ M were collected successively to assess the sensitivity of these plasmonic nanogaps as SERS probes, as shown in Figure 4e. Remarkably, the characteristic peaks of R6G from most nanogaps remains detectable even at an extremely low concentration of $10^{-12}$ M (Figure 4f).

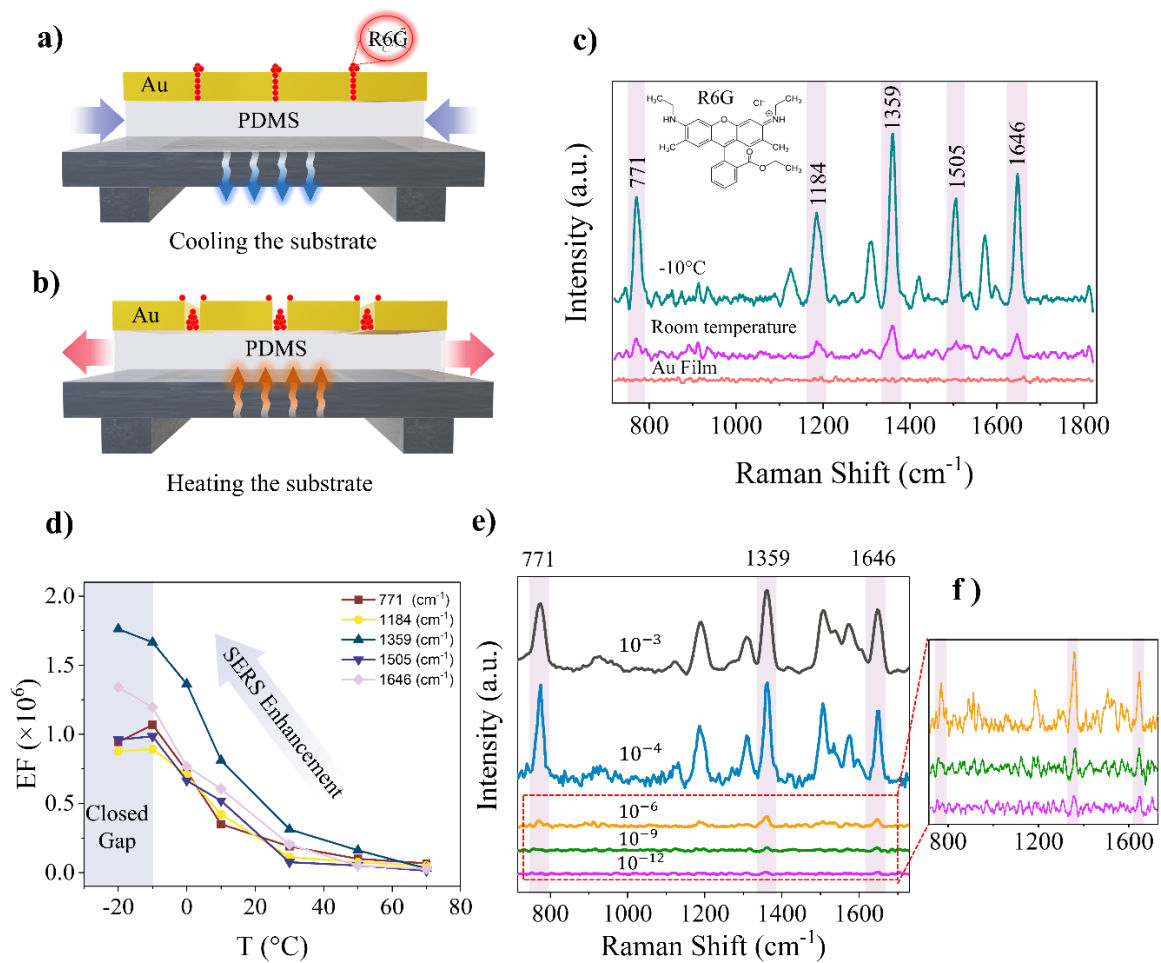

**Figure 4.** a) and b) Schematic representations of the nanogap under cooling and heating conditions, respectively, highlighting the advantage of precise positioning of the molecules within the hotspot region. c) SERS spectra of molecules measured on an Au thin film without the nanogap, nanogap pattern at room temperature, and at -10°C. d) Enhancement factor of Raman intensity for various characteristic peaks, extracted from Supporting Information Figure S8a. e) SERS spectra of R6G at different concentrations and f) an enlarged view of the spectra at low concentrations.

We evaluated the reproducibility of our nanogap-based SERS structure by comparing the SERS signals of two different samples. The SERS spectra were recorded during cooling and heating with $10^{-4}$ M R6G of sample A (see **Figure S11**a and S11b in Supporting Information) and sample B (see Figure S11c and S11d in Supporting Information), demonstrating the excellent modulation of the Raman shifts. To further assess the reproducibility, Raman spectra were recorded at 10 randomly selected points within a range of 1 cm$^{-1}$ along the nanogap, as shown in **Figure S12** in Supporting Information. The relative standard deviation of the Raman signal was calculated to be 1.99 %, indicating a high degree of consistency.

## 3.Conclusion

Our research shows a significant breakthrough in the development of thermally tunable nanogaps for SERS applications. By combining flexible substrates with advanced nanofabrication techniques, we have achieved precise control over the dimensions of the nanogaps, enabling dynamic modulation of SERS signals. The tunable nanogaps improve sensitivity and enable effective detection of molecules with an amplification factor suitable for single molecule analysis. With a detection limit as low as $10^{-12}$ M, our approach offers high sensitivity and adaptability, making it a powerful tool for various applications, including chemical sensing and biological detection. This work lays the foundation for the integration of tunable nanogaps into broader sensing technologies and expands their potential in areas such as environmental monitoring and medical diagnostics.

## 4. Experimental Section

*Molecule Preparation:* R6G powders were dissolved in ethanol to prepare solutions of different concentrations. The samples were functionalized by immersing them in these solutions for a certain time. After immersion, the samples were removed, rinsed with ethanol to remove unbound molecules, and then dried with $N_2$.

*Microwave Setup*: Microwave measurements have been conducted using Agilent Technologies E5063A ENA series network analyzer, configured with a vertical waveguide assembly operating in Ku-band frequency range (12–18 GHz). For the measurements, a set of two open rectangular waveguides (62EWGN) connected to the network analyzer is used and the sample is placed in the middle of this setup. The aperture size to support the TE10 mode for Ku-band frequency is 15.80 × 7.90 mm.

*Temperature setup:* To investigate the temperature-dependent transmission characteristics, a laboratory-installed Peltier setup is employed in the microwave setup. The sample is placed on a conductive plate connected to the Peltier device to ensure uniform heat distribution. It is important that the sample is free to expand or contract as no adhesive tapes were used to restrict its movement. A temperature sensor connected to a monitoring system was positioned close to the sample to enable precise temperature measurements in real time.

*Electrical Connections*: The electrical conductivity of the sample was measured as a function of temperature using a Keithley 2450 source meter connected to the sample via copper wires. A constant input voltage of 1V was applied, and the resulting changes in the output current were recorded.

*Theoretical Analysis*: The EM field distribution in metallic nanogaps is simulated using two-dimensional (2D) finite element analysis (FEA) through the COMSOL Multiphysics 6.2 software. The simulations employed a plane wave source with an electric field polarized along the x-axis, propagating along the y-axis. The nanogap width has varied from 5 to 100 nm, while the depth ranged from 30 to 200 nm. To avoid the tip-enhancement effects of the EM field, rounded corners with a 2.5 nm radius are introduced at the top edges of the nanogap. Perfectly matched layers (PML) and scattering boundary conditions (SBC) have been used to minimize reflections at the simulation boundaries. The simulation covered a volume of approximately 2.5 × 1.5 μm, which represented half of the spatial period and three times the wavelength of the incident light. Due to the presence of two mirror planes, only 1/4 of the original simulation volume was required.

*SERS Measurements:* SERS measurements were performed using a confocal Raman spectrometer (WITec, Alpha300R, Germany) with a 532 nm laser as excitation source. The laser was focused on the sample surface through a 50× objective lens (NA = 0.5), with a spot size of approximately 1 μm. The laser power on the sample surface was limited to 5 mW, with an exposure time of 1 second. Each Raman spectrum was recorded from a single nanogap, with repeatability confirmed by 10 accumulations. For the temperature-dependent SERS measurements, the sample was mounted on a temperature-controlled stage (THMS600). This setup was connected to a commercially available temperature controller (TMS94) and a supply of liquid nitrogen.

## Supporting Information

Supporting Information is available from the Wiley Online Library or from the author.

## Acknowledgements

This work was supported by the National Research Foundation of Korea (NRF) grant funded by the Korean government (MSIT: NRF-2015R1A3A2031768), the National R&D Program through the National Research Foundation of Korea (NRF) funded by Ministry of Science and ICT (2022M3H4A1A04096465), the Basic Science Research Program through the National Research Foundation of Korea (NRF) funded by the Ministry of Education (NRF-2022R1I1A1A01073838), and the MSIT (Ministry of Science and ICT), Korea, under the ITRC (Information Technology Research Center) support program (IITP-2023-RS-2023-00259676) supervised by the IITP (Institute for Information & Communications Technology Planning & Evaluation).

## Conflict of Interest

The authors declare no conflict of interest.

## Author Contributions

Mahsa Haddadi Moghaddam. conceived the idea, fabricated the samples, designed, and performed the experiments, the data analysis and wrote the manuscript; Sobhagyam Sharma. helped the fabrication and performed the experiments and data analysis; Daehwan Park conducted theoretical analysis, Dai Sik Kim. led the work, reviewed, edited, and supervised this work.

*Mahsa Haddadi Moghaddam and Sobhagyam Sharma contributed equally.

## Data Availability Statement

The data that support the findings of this study are available in the supplementary material of this article.

## References

[1] A. Das, U. Pant, C. Cao, R. S. Moirangthem, H. B. Kamble, *ACS Applied Optical Materials* **2023**, 1, 1938.

[2] W. Fang, S. Jia, J. Chao, L. Wang, X. Duan, H. Liu, Q. Li, X. Zuo, L. Wang, L. Wang, N. Liu, C. Fan, *Science Advances*, 5, eaau4506.

[3] P. Gu, W. Zhang, G. Zhang, *Advanced Materials Interfaces* **2018**, 5, 1800648.

[4] T. Itoh, M. Procházka, Z.-C. Dong, W. Ji, Y. S. Yamamoto, Y. Zhang, Y. Ozaki, *Chemical Reviews* **2023**, 123, 1552.

[5] Y. Ma, D. Sikdar, A. Fedosyuk, L. Velleman, D. J. Klemme, S.-H. Oh, A. R. J. Kucernak, A. A. Kornyshev, J. B. Edel, *ACS Nano* **2020**, 14, 328.

[6]     A. D. McFarland, M. A. Young, J. A. Dieringer, R. P. Van Duyne, *The Journal of Physical Chemistry B* **2005**, 109, 11279.

[7]     H. Xu, E. J. Bjerneld, M. Käll, L. Börjesson, *Physical Review Letters* **1999**, 83, 4357.

[8]     X. Zhao, X. Liu, D. Chen, G. Shi, G. Li, X. Tang, X. Zhu, M. Li, L. Yao, Y. Wei, W. Song, Z. Sun, X. Fan, Z. Zhou, T. Qiu, Q. Hao, *Nature Communications* **2024**, 15, 5855.

[9]     Z. Zuo, S. Zhang, Y. Wang, Y. Guo, L. Sun, K. Li, G. Cui, *Nanoscale* **2019**, 11, 17913.

[10]    P. K. Jain, W. Huang, M. A. El-Sayed, *Nano Letters* **2007**, 7, 2080.

[11]    P. Sadeghi, K. Wu, T. Rindzevicius, A. Boisen, S. Schmid,  **2018**, 7, 497.

[12]    Y.-H. Wang, S. Zheng, W.-M. Yang, R.-Y. Zhou, Q.-F. He, P. Radjenovic, J.-C. Dong, S. Li, J. Zheng, Z.-L. Yang, G. Attard, F. Pan, Z.-Q. Tian, J.-F. Li, *Nature* **2021**, 600, 81.

[13]    J. Xu, X. Zhu, S. Tan, Y. Zhang, B. Li, Y. Tian, H. Shan, X. Cui, A. Zhao, Z. Dong, J. Yang, Y. Luo, B. Wang, J. G. Hou, *Science* **2021**, 371, 818.

[14]    M. Charconnet, C. Kuttner, J. Plou, J. L. García-Pomar, A. Mihi, L. M. Liz-Marzán, A. Seifert, *Small Methods* **2021**, 5, 2100453.

[15]    Y. Fang, N.-H. Seong, D. D. Dlott, *Science* **2008**, 321, 388.

[16]    F. Huang, J. J. Baumberg, *Nano Letters* **2010**, 10, 1787.

[17]    D. Jonker, K. Srivastava, M. Lafuente, A. Susarrey-Arce, W. van der Stam, A. van den Berg, M. Odijk, H. J. G. E. Gardeniers, *ACS Applied Nano Materials* **2023**, 6, 9657.

[18]    Y.-F. Chou Chau, T. Y. Ming, C.-T. Chou Chao, R. Thotagamuge, M. R. R. Kooh, H. J. Huang, C. M. Lim, H.-P. Chiang, *Scientific Reports* **2021**, 11, 18515.

[19]    H. Le-The, J. J. A. Lozeman, M. Lafuente, P. Muñoz, J. G. Bomer, H. Duy-Tong, E. Berenschot, A. van den Berg, N. R. Tas, M. Odijk, J. C. T. Eijkel, *Nanoscale* **2019**, 11, 12152.

[20]    H. Liang, L. Jiang, H. Li, J. Zhang, Y. Zhuo, R. Yuan, X. Yang, *ACS Sensors* **2023**, 8, 1192.

[21]    J. M. McMahon, S. Li, L. K. Ausman, G. C. Schatz, *The Journal of Physical Chemistry C* **2012**, 116, 1627.

[22]    R. Pan, Y. Yang, Y. Wang, S. Li, Z. Liu, Y. Su, B. Quan, Y. Li, C. Gu, J. Li, *Nanoscale* **2018**, 10, 3171.

[23]    S. Yan, H. Tang, J. Sun, C. Zhu, Q. Pan, B. Chen, G. Meng, *Advanced Optical Materials* **2024**, 12, 2302010.

[24]    P. Zhan, T. Wen, Z.-g. Wang, Y. He, J. Shi, T. Wang, X. Liu, G. Lu, B. Ding, *Angewandte Chemie International Edition* **2018**, 57, 2846.

[25]    Q. Zhao, H. Yang, B. Nie, Y. Luo, J. Shao, G. Li, *ACS Applied Materials & Interfaces* **2022**, 14, 3580.

[26]    Y. Ye, J. Wang, Z. Fang, Y. Yan, Y. Geng, *ACS Applied Materials & Interfaces* **2024**, 16, 10450.

[27]    H. Im, K. C. Bantz, N. C. Lindquist, C. L. Haynes, S.-H. Oh, *Nano Letters* **2010**, 10, 2231.

[28]    K. D. Alexander, K. Skinner, S. Zhang, H. Wei, R. Lopez, *Nano Letters* **2010**, 10, 4488.

[29]    R. E. Darienzo, O. Chen, M. Sullivan, T. Mironava, R. Tannenbaum, *Materials Chemistry and Physics* **2020**, 240, 122143.

[30]    J. Jeong, H. W. Kim, D.-S. Kim,  **2022**, 11, 1231.

[31]    M. Li, Y. Liu, X. Liu, Y. Zhang, T. Zhu, C. Feng, Y. Zhao, *Spectrochimica Acta Part A: Molecular and Biomolecular Spectroscopy* **2022**, 275, 121159.

[32]    J. Lin, J. Yu, O. U. Akakuru, X. Wang, B. Yuan, T. Chen, L. Guo, A. Wu, *Chemical Science* **2020**, 11, 9414.

[33]    M. H. Moghaddam, D. J. C. Dalayoan, D. Park, Z. Wang, H. Kim, S. Im, K. Ji, D. Kang, B. Das, D. S. Kim, *ACS Photonics* **2024**, 11, 3239.

[34]    Y. Qi, V. Brasiliense, T. W. Ueltschi, J. E. Park, M. R. Wasielewski, G. C. Schatz, R. P. Van Duyne, *Journal of the American Chemical Society* **2020**, 142, 13120.

[35]    T. Köker, N. Tang, C. Tian, W. Zhang, X. Wang, R. Martel, F. Pinaud, *Nature Communications* **2018**, 9, 607.

[36]    Y. Liu, L. Zhang, X. Liu, Y. Zhang, Y. Yan, Y. Zhao, *Spectrochimica Acta Part A: Molecular and Biomolecular Spectroscopy* **2022**, 270, 120803.

[37]    S. Nie, S. R. Emory, *Science* **1997**, 275, 1102.

[38]    S. Kim, B. Das, K. H. Ji, M. H. Moghaddam, C. Chen, J. Cha, S. Namgung, D. Lee, D.-S. Kim,  **2023**, 12, 1481.

[39]    H. S. Yun, J. Jeong, D. Kim, D. S. Kim, presented at *2018 43rd International Conference on Infrared, Millimeter, and Terahertz Waves (IRMMW-THz)*, 9-14 Sept. 2018, **2018**.

[40]    P. Verma, *Chemical Reviews* **2017**, 117, 6447.

[41]    A. Bamido, N. Shettigar, A. Thyagrajan, D. Banerjee, presented at *2021 20th IEEE Intersociety Conference on Thermal and Thermomechanical Phenomena in Electronic Systems (iTherm)*, 1-4 June 2021, **2021**.

[42] Y. Ni, J. Huang, S. Li, X. Dong, T. Zhu, W. Cai, Z. Chen, Y. Lai, *ACS Applied Materials & Interfaces* **2021**, 13, 53271.

[43] L. Wu, J. Qian, J. Peng, K. Wang, Z. Liu, T. Ma, Y. Zhou, G. Wang, S. Ye, *Journal of Materials Science: Materials in Electronics* **2019**, 30, 9593.

[44] F. C. Nix, D. MacNair, *Physical Review* **1941**, 60, 597.

[45] Z. Wang, A. A. Volinsky, N. D. Gallant, *Journal of Applied Polymer Science* **2014**, 131.

[46] Y. Shi, M. Hu, Y. Xing, Y. Li, *Materials & Design* **2020**, 185, 108219.

[47] L. Cui, S. An, H. Yit Loong Lee, G.-X. Liu, H. Wang, H.-Y. Wang, L. Wu, Z. Dong, L. Wang, *Nano Letters* **2024**, 24, 9337.

[48] S. X. Leong, Y. X. Leong, E. X. Tan, H. Y. F. Sim, C. S. L. Koh, Y. H. Lee, C. Chong, L. S. Ng, J. R. T. Chen, D. W. C. Pang, L. B. T. Nguyen, S. K. Boong, X. Han, Y.-C. Kao, Y. H. Chua, G. C. Phan-Quang, I. Y. Phang, H. K. Lee, M. Y. Abdad, N. S. Tan, X. Y. Ling, *ACS Nano* **2022**, 16, 2629.

[49] J.-M. Nam, J.-W. Oh, H. Lee, Y. D. Suh, *Accounts of Chemical Research* **2016**, 49, 2746.